\documentstyle[epsfig,11pt]{article}
\begin{document}
\vskip 0.5cm
\begin{center}
{\Large{SEARCH FOR MASSIVE RARE PARTICLES\\
\vskip 0.3cm
 WITH MACRO}}
\end{center}

\vskip .7 cm
\begin{center}

{\bf The MACRO Collaboration} \\
\nobreak\bigskip\nobreak
\pretolerance=10000
M.~Ambrosio$^{12}$, 
R.~Antolini$^{7}$, 
G.~Auriemma$^{14,a}$, 
D.~Bakari$^{2,17}$,
A.~Baldini$^{13}$, 
G.~C.~Barbarino$^{12}$, 
B.~C.~Barish$^{4}$, 
G.~Battistoni$^{6,b}$, 
R.~Bellotti$^{1}$, 
C.~Bemporad$^{13}$
% ...
\\
\vspace{1.5 cm}
\footnotesize
1. Dipartimento di Fisica dell'Universit\`a di Bari and INFN, 70126 
Bari,  Italy \\
2. Dipartimento di Fisica dell'Universit\`a di Bologna and INFN, 
 40126 Bologna, Italy \\
3. Physics Department, Boston University, Boston, MA 02215, 
USA \\
4. California Institute of Technology, Pasadena, CA 91125, 
USA \\
5. Department of Physics, Drexel University, Philadelphia, 
PA 19104, USA \\
6. Laboratori Nazionali di Frascati dell'INFN, 00044 Frascati (Roma), 
Italy \\
7. Laboratori Nazionali del Gran Sasso dell'INFN, 67010 Assergi 
(L'Aquila),  Italy \\
8. Depts. of Physics and of Astronomy, Indiana University, 
Bloomington, IN 47405, USA \\
9. Dipartimento di Fisica dell'Universit\`a dell'Aquila  and INFN, 
 67100 L'Aquila,  Italy \\
10. Dipartimento di Fisica dell'Universit\`a di Lecce and INFN, 
 73100 Lecce,  Italy \\
11. Department of Physics, University of Michigan, Ann Arbor, 
MI 48109, USA \\	
12. Dipartimento di Fisica dell'Universit\`a di Napoli and INFN, 
 80125 Napoli,  Italy \\	
13. Dipartimento di Fisica dell'Universit\`a di Pisa and INFN, 
56010 Pisa,  Italy \\	
14. Dipartimento di Fisica dell'Universit\`a di Roma ``La Sapienza" and INFN, 
 00185 Roma,   Italy \\ 	
15. Physics Department, Texas A\&M University, College Station, 
TX 77843, USA \\	
16. Dipartimento di Fisica Sperimentale dell'Universit\`a di Torino and INFN,
 10125 Torino,  Italy \\	
17. L.P.T.P., Faculty of Sciences, University Mohamed I, B.P. 524 Oujda, Morocco \\
18. Dipartimento di Fisica dell'Universit\`a di Roma Tre and INFN Sezione Roma Tre, 
 00146 Roma,   Italy \\ 	
$a$ Also Universit\`a della Basilicata, 85100 Potenza,  Italy \\
$b$ Also INFN Milano, 20133 Milano, Italy\\
$c$ Also Istituto TESRE/CNR, 40129 Bologna, Italy \\
$d$ Also Scuola Normale Superiore di Pisa, 56010 Pisa, Italy\\
$e$ Also Universit\`a di Trieste and INFN, 34100 Trieste, 
Italy \\
$f$ Also Dipartimento di Energetica, Universit\`a di Roma, 
 00185 Roma,  Italy \\
$g$ Also Institute for Nuclear Research, Russian Academy
of Science, 117312 Moscow, Russia \\
$h$ Also Institute for Space Sciences, 76900 Bucharest, Romania \\
$i$ The Colorado College, Colorado Springs, CO 80903, USA\\
$l$ Also Department of Physics, Pennsylvania State University, 
University Park, PA 16801, USA\\
$m$ Also Dipartimento di Fisica dell'Universit\`a della Calabria, Rende (Cosenza), 
Italy \\ 	
$n$ Also Department of Physics, James Madison University, Harrisonburg, 
VA 22807, USA\\
$p$ Also RPD, PINSTECH, P.O. Nilore, Islamabad, Pakistan\\

\end{center}

\vskip 1.5 cm

\begin{abstract}

Massive rare particles have been searched for in the penetrating cosmic radiation 
using the MACRO apparatus at the Gran Sasso National Laboratories.
Liquid scintillators, streamer tubes and nuclear track detectors have been
 used to search for magnetic monopoles (MMs). 
 Based on no observation of such signals, stringent flux limits  
are established for MMs as slow as a few $10^{-5}c$. 
 
The methods based on the scintillator and on the nuclear 
track subdetectors were also applied to search for nuclearites. \par
Preliminary results of the searches for charged Q-balls are also presented.  

%
%

\end{abstract}

\section{INTRODUCTION}

 One of the primary aims of the MACRO  experiment at the 
Gran Sasso underground Laboratories 
is the search for magnetic monopoles at the mass scale of  
Grand Unified Theories (GUTs) of the electroweak and strong interactions
 \cite{Preskill79} with a sensitivity  
well below the
Parker bound ($10^{-15}$ cm$^{-2}$ s$^{-1}$ sr$^{-1}$) \cite{Turner82}
in the velocity range $4 \cdot 10^{-5}<\beta<1$, $\beta=v/c$. \par

 MACRO has three subdetectors: liquid scintillation counters,
limited streamer tubes
and nuclear track detectors (CR39 and Lexan) arranged in a modular
structure of six ``supermodules" (SM's).
Each SM is divided into a lower and an upper (``Attico") part and comes with
separate mechanical structure and electronics readout. 
The full detector has global dimensions of $76.5 \times 12 \times 9.3$
m$^3$ \cite{Ahlen93} and provides a total acceptance to an isotropic flux of 
particles of $\sim 10,000$ m$^2$sr.
The detector has been built and equipped with electronics during the years
1988-1995. Data taking began in 1989 with the first SM; since the fall of 
1995 it is running in its final
configuration.  
The response to slow and
fast particles of the scintillators, streamer tubes and nuclear track 
detectors  was experimentally studied
 \cite{Ahlen83,Batt88,Cecco96}.
The three subdetectors ensure redundancy of information, cross-checks
and independent signatures for possible MM candidates. \par
The analyses presented here, based on the various subdetectors in a
stand-alone and in a combined way,
refer to direct detection of bare MMs of one
 unit Dirac
 charge ($g_D=137/2e$),  catalysis cross section $\sigma_{cat} <1$ mb
and  isotropic flux (we consider MMs with enough kinetic energy to
traverse the Earth); this last condition sets a $\beta$ dependent mass
threshold ($\sim 10^{17}$ GeV for $\beta \sim 5 \cdot 10^{-5}$, and
lower for faster MMs). Since no MM candidate  was found we quote new 
flux upper limits at the level of 
$2.5 \cdot  10^{-16}$ cm$^{-2}$ s$^{-1}$ sr$^{-1}$ for 
$\beta > 5 \cdot 10^{-5}$.
 \par

``Strange Quark Matter" (SQM) should consist of aggregates of 
comparable amounts of $u$,
$d$ and $s$ quarks; it might be the ground state of QCD \cite{Witten}.
If bags of SQM were produced in a first-order phase transition in the early
universe, they  could be candidates for the Dark Matter (DM), and might be found
in the cosmic radiation reaching the Earth. SQM in the cosmic
radiation is commonly known as ``nuclearite" and ``strangelet"\cite{Rujula-Glas84}.

Q-balls should be  aggregates of squarks, sleptons and Higgs fields 
\cite{Coleman,Kusenko}. 
They could have been produced in the early Universe, and
 may contribute to the Cold Dark Matter. Heavy Q-balls may have originated 
in the course
of a phase transition or they could have been produced via fusion processes,
reminiscent of the big bang nucleosynthesis. Small Q-balls could also be
pair-produced in very high energy collisions.
Relic Q-balls can be separated in two classes: Supersymmetric
Electrically Charged Solitons (SECS) and Supersymmetric  Electrically 
Neutral Solitons (SENS).

Some of the methods used for the MM searches may also be applied to search
for nuclearites and for charged Q-balls (SECS). We quote upper limits
for $\beta > 5\cdot 10^{-5}$.

\section{SEARCHES FOR MAGNETIC MONOPOLES}

A flux of cosmic GUT supermassive magnetic monopoles may reach the Earth.
 The velocity
spectrum of these
MMs  could be in the range $4\cdot 10^{-5} <\beta <0.1$.
Our searches for MMs exploit their energy loss mechanisms  in each of the 
three MACRO subdetectors.  

In scintillators the fraction of energy loss which is effective for the 
detection is the
excitation energy loss which leads to the emission of light; in streamer
tubes it is the ionization 
energy loss in the gas; in nuclear track
detectors it is
the Restricted Energy Loss (REL), i.~e., the energy deposited within $\sim$
10 nm from
the MM trajectory. In Ref.~\cite{Derkaoui99} a thorough analysis is made 
of these losses and
of their dependence on the MM velocity.

Independent and combined monopole analyses were performed using the 
scintillator, streamer tube and nuclear track subdetectors in 
different ranges of velocity.
As already stated the results presented here apply to bare $g_D$ MMs and 
$\sigma_{cat} <1$ mb; for the streamer tube analysis the 
dependence of the results on $\sigma_{cat}$ is discussed below.

\subsection{Searches with scintillators}

 The searches with
the liquid scintillator subdetector use different
specialized triggers covering specific velocity regions; the
searches are grouped into searches for low velocity
($10^{-4} < \beta < 10^{-3}$),
 medium velocity ($10^{-3} < \beta < 10^{-1}$) and high velocity
($\beta>0.1)$ particles.

\subsubsection{Low velocity monopole searches}

Previous searches using  data collected with the Slow Monopole Trigger (SMT)
and  Waveform Digitizer (WFD)
were reported  in Ref.~\cite{Ambrosio97}, see curves ``A'', ``B'' in
Fig.~1.
 A new custom made 200 MHz WFD system was implemented in 1995 
which  improves by at
least a factor of two the sensitivity
to very slow monopoles ($\beta\sim10^{-4}$) and by over a factor of five
the sensitivity to relativistic monopoles with
respect to previous conditions. The sensitivity
of the SMT/WFD  was tested with  LED pulses, of
$\sim6.3$ $\mu$s duration, corresponding to $\beta\sim10^{-4}$,
down to the level of few tens of
single photoelectrons, which is
the signature of a slow monopole.
A waveform analysis procedure consisted in scanning off-line the corresponding wave forms
and in software simulation of the function of both the analog
and digital part of the SMT circuitry on an event-by-event basis.
We plan to report on this search in the near future.

\subsubsection{Medium and high velocity monopole searches}

The data collected by the PHRASE (Pulse Height Recorder and Synchronous
Encoder) trigger
 are used to search for MMs in the
range $1.2 \cdot 10^{-3} <\beta < 10^{-1}$ \cite{Ambrosio97,Ambrosio92}.
The events
are selected requiring hits in a maximum of four adjacent scintillation 
counters, with a
minimum energy deposition of 10~MeV in two different scintillator layers.
Events with $1.2 \cdot 10^{-3} < \beta < 5 \cdot 10^{-3}$
   are rejected because their
pulse width is smaller than the expected counter crossing time; events
with $5 \cdot 10^{-3} < \beta <
10^{-1}$
 are rejected because the light produced is much lower than
that expected for a MM.
 The analyses refer to  data
collected by
the MACRO lower part from October 1989 to the end of 1999 and also by the
 Attico from June 1995 to the end of 1999.
No candidate survives;  the
 $90\%$ C.L. flux upper limit is $2.6 \cdot 10^{-16}$~
cm$^{-2}$~s$^{-1}$~sr$^{-1}$ (curve ``D'' in
Fig.~1).\par
A previous search for MMs with $\beta >10^{-1}$ based on the ERP
(Energy Reconstruction Processor) trigger \cite{Ambrosio97,Ambrosio92}
is  included in Fig.~1 ( curve ``C'').

\subsection{Search using the streamer tubes}
The streamer tube search was described in
Refs.~\cite{Ambrosio97,Ambrosio95}.
 The detection
of MMs of $10^{-4} < \beta < 10^{-3}$ is based on the 
Drell and Penning effects in 
the gas mixture ($73\%$ He and $27\%$ n-pentane)
filling the tubes \cite{Ambrosio95,Drell}.
The analysis is based on the search for single tracks in the streamer
tubes  and on the measurement
of the velocity with the ``time track''. Only the horizontal
streamer planes of the lower MACRO structure are used in the
trigger; the Attico and the  vertical planes
are used for event  reconstruction.
Data were collected from January 1992 to end of 1999
for a live-time of $6.6 \cdot 10^{4}$ hours.
The trigger and the analysis chain were checked to be velocity independent.
The global
efficiency was  estimated by computing the ratio of the rate of single
muons reconstructed by this analysis to the expected one \cite{Ambrosio95}.
The overall efficiency was $74\%$. The detector acceptance, computed
by a Monte Carlo simulation
 including  geometrical and trigger requirements,
is 4250 m$^2$ sr . No monopole candidate was found.
For $1.1 \cdot 10^{-4} < \beta < 5 \cdot 10^{-3}$  the flux upper limit is
$ 3.1 \cdot 10^{-16}$ cm$^{-2}$ s$^{-1}$ sr$^{-1}$
at 90\% C.L. (Fig.~1, curve ``Streamer'').


\subsubsection{ Catalysis of nucleon decay}
\par Detailed Monte Carlo simulations were performed to study the effects of 
nucleon decay catalyzed by a GUT magnetic monopole on the streamer monopole
trigger and its effects on the relative analysis.
Both the physical process and the detector response were introduced in the 
code, taking into account the theoretical predictions on the
cross section and on the decay channels. 
\par Many samples (each of 10,000 monopole events) were generated with constant
monopole velocity $\beta = 10^{-4}$, $5\cdot 10^{-3}$, $10^{-3}$, $5\cdot
10^{-2}$ and $10^{-2}$, and with catalysis cross sections
 $\sigma_{cat} = 10^{-26}$, $10^{-25}$, $5\cdot 10^{-25}$ and
$10^{-24}$~cm$^2$ (in the last two cases other samples with 
$\beta = 2\cdot 10^{-4}$ and $2\cdot 10^{-3}$ were produced); also a
sample with no catalysis was simulated as a term of reference. These simulations were
performed according to two different theoretical models for the catalysis 
cross section, one which considers it to have the same $\beta$ dependence 
for both
protons and neutrons and a second one which assumes it to be 
$\beta$--enhanced in case of
protons.
\par All  samples were analyzed with the same program used for the
real data. This allowed a study of  the detection efficiency 
as a function of $\beta$ and $\sigma_{cat}$. As a consequence
new upper limits can be established which take into account also this process.
Figs.~2 and 3 show for a catalysis event with
$\beta = 10^{-3}$ and $\sigma_{cat} = 10^{-25}~$cm$^2$
the time and wire views, respectively: in the time view 
the monopole
straight track and the catalysis hits are clearly distinguishable. Fig.~4
shows the distributions of the reconstructed $\beta$: 
the distributions are exactly peaked on the input values, which means that the
reconstruction code works well also in the presence of catalysis hits. Finally
Fig.~5 shows the upper limits vs $\beta$ for different $\sigma_{cat}$: 
for low catalysis cross
sections ($\sigma_{cat} \le 10^{-25}$~cm$^2$) the difference is negligible,
 while for
higher values it becomes more important.
\par A new analysis is in progress searching for catalysis events in
the real data. Moreover checks are being carried out to see how the catalysis
may affect the combined fast monopole analysis (Sect. 2.4).
\par

\par 

\par

\subsection{Search using the nuclear track subdetector}

The nuclear track subdetector
covers a surface of 1263 m$^{2}$ and the acceptance
for fast MMs is 7100 m$^2$ sr.
 The subdetector is used as a stand-alone detector
and in a ``triggered mode" by
the scintillator and streamer tube systems.
A detailed description of the method of searching for MMs is
given in Ref.~\cite{Giorgini}.
On May 2000 we began the massive etching of the CR39 sheets using the
Bologna and Gran Sasso facilities, at the rate of about 40 m$^2$/month.
 An area of 368 m$^2$ of CR39 has been analysed,
with an average exposure time of 8.5 years.
No candidate was found; the  90$\%$
C.L. upper limits on the MM flux are at the level of
$3.7 \cdot 10^{-16}$ cm$^{-2}$ s$^{-1}$ sr$^{-1}$ at $\beta\sim$ 1,
and $5.4 \cdot 10^{-16}$ cm$^{-2}$ s$^{-1}$ sr$^{-1}$ at $\beta\sim 10^{-4}$
 (Fig.~1, curves ``CR39'').

\subsection{Combined searches for fast monopoles}

A search for fast MMs with scintillator or streamer tubes
is affected by the background due
to energetic muons with large energy losses
(the nuclear track detector is not affected).
Two analyses, which combine the use of the three subdetector systems, were 
performed in order to 
achieve the highest rejection imposing looser requirements.

\subsubsection{Streamer tubes+ERP}
The analysis procedure is based on the scintillator and streamer tube data; the
nuclear track detector is used as a final tool for rejection/confirmation of the
selected candidates.
The trigger requires at least one fired scintillation counter
and 7 hits in the  horizontal streamer planes.
Candidates are selected on the basis of the scintillator light
yield and  of the digital (tracking) and
analog (pulse charge) information from the streamer tubes.
A further selection is then applied on the streamer tube pulse charge.
After corrections for gain variations,  geometrical
 and electronic non-linear effects \cite{Batt97b},
a $90\%$ efficiency cut is applied on the average streamer charge.
 Possible candidates ($\sim$~2/year) are analysed in the
corresponding nuclear track detector modules. The analysis
refers to about 36,980 live hours
 with an average efficiency of $77\%$.
The geometrical acceptance, computed by Monte Carlo
methods, including the analysis requirements, is 3565 m$^2$ sr.
No candidate survives; the $90\%$ C.L. flux upper limit is
$6.3 \cdot 10^{-16}$ cm$^{-2}$ s$^{-1}$ sr$^{-1}$ for MMs with
$5 \cdot 10^{-3} < \beta < 0.99$ (curve ``E'' in Fig.~1).\par

\subsubsection{PHRASE+Streamer tubes} 

MMs with $\beta>10^{-2}$ are searched for by combining
the streamer tube and  PHRASE triggers. Streamer tubes are
used to reconstruct the trajectory and pathlength, scintillators are used
to measure the velocity and the light yield.  Selected events ($\sim 50$/year)
 have
a single track and an energy deposition
$>$~200~MeV in three scintillator layers.
 The  event energy loss
is compared to that expected for a
monopole with the same velocity. The analysis refers to about $8528$ live hours
from May, $1997$ to June, $1998$. No candidate survives.
The geometrical acceptance, including  analysis cuts, is
3800 m$^2$ sr.  The $90\%$ C.L  flux upper limit is $2.3 \cdot
10^{-15}$ cm$^{-2}$ s$^{-1}$ sr$^{-1}$ (curve ``F'' in
Fig.~1).

\section{SEARCHES FOR NUCLEARITES}

The main energy loss mechanism for nuclearites passing through matter is
elastic or quasi-elastic collisions \cite{Rujula-Glas84}:
$$ \frac{dE}{dx} = \sigma \rho v^2 $$
where $\sigma$ is the nuclearite cross section, $v$ its velocity and
$\rho$ the mass density of the traversed medium.

For nuclearites with masses $M \geq 8.4 \cdot 10^{14}$~GeV ($\simeq
1.5$~ng) the cross section may be approximated as:
$$ \sigma \simeq \pi \cdot \left( \frac{3M}{4 \pi \rho_N} \right)^{3/2}$$
where $\rho_N$ (the density of SQM) is estimated to be $\rho_N \simeq
3.5 \cdot 10^{14}$ g/cm$^3$ \cite{Chin71}. For lighter nuclearites
the collisions are governed by their electronic clouds, yielding $\sigma
\simeq \pi \cdot 10^{-16}$~cm$^2$.

Assuming galactic velocities, $\beta \simeq 2 \cdot 10^{-3}$,
 nuclearites with masses $\leq 5 \cdot 10^{11}$ GeV cannot reach
the detector; for $ 5 \cdot 10^{12} \leq M \leq 10^{21}$ GeV only downward
going nuclearites can reach it; for  $M > 10^{22}$ GeV nuclearites can
reach MACRO from all
directions.
Scintillators are sensitive to the blackbody radiation emitted along
the heated nuclearite paths down to
$\beta \simeq 5 \cdot 10^{-5}$. The CR39 is sensitive to nuclearites
 down to $\beta \sim 10^{-5}$ \cite{Ambrosio99}. The density of the gas
mixture in the streamer tubes is too low to produce energy losses
yielding ionization, so the streamer tubes are not useful for
nuclearite searches.

Individual flux limits for nuclearites from the scintillator and CR39
subdetectors are presented in
Fig.~6; curves ``a - d''  refer to earlier searches with scintillators
 \cite{Ambrosio99}; curves  ``e'' and ``f'' are the updated limits 
obtained using
the PHRASE system (Sect.~2.1.2) and CR39 nuclear track detectors, 
respectively.\par

\section{SEARCHES FOR CHARGED Q-BALLS}
The detection of Q-balls in MACRO is discussed in Ref.~\cite{Ouchrif}.
In that work it is assumed that 
the main
contribution to the  energy losses of SECS passing through matter with
velocities in the range
$10^{-4}<\beta<10^{-2}$ are  due to the interaction of the SECS
positive charge with the nuclei (nuclear contribution) and
with the electrons (electronic contribution) of the traversed medium.
Other energy loss mechanisms, like nuclearite energy loss, could be considered
and are under investigation.\par
The MACRO subdetectors are sensitive to SECS for any value of
the electric charge $Z_Q \geq 1$.        
The observational signatures of SECS are characterized
by substantial energy release along a straight track with no attenuation
throughout the detector \cite{Kusenko}. The energy losses of SECS in different 
subdetectors are given in Ref.~\cite{Ouchrif}. CR39 and scintillators are 
sensitive to Q-balls with $Z_Q \geq 1$ for $3\cdot 10^{-5} \le \beta \le
 0.1 $ and
$\beta \ge 6\cdot 10^{-5}$, respectively; in streamer tubes the detection
threshold is at $\beta=2\cdot 10^{-3}$.
We checked that the methods to search for MMs based on the PHRASE system 
(Sect.~2.1.2 ), on the  
streamer tube system 
(Sect.~2.2) and on the nuclear track detector (Sect.~2.3) 
can be applied to SECS. Flux limits for Q-balls with $Z_Q \geq 1$  
obtained with the MACRO scintillators (curves ``PHRASE'') and
with streamer tubes (curve ``Streamer'')  are shown in Fig.~7; 
the limit obtained
with CR39 (curve ``CR39'' in Fig.~7) applies to Q-balls with $Z_Q=1$.

\section{CONCLUSIONS}

No MM, no nuclearite, no Q-ball candidates were found in any of 
these searches.\par
 The 90\% C.L. flux limits for MMs versus $\beta$ are shown in Fig.~1.
 The global MACRO limit is computed as $2.3 / X_{total}$ where
$X_{total}= \sum_{i}^{} X'_i $,
and the $X'_i$  are the independent time integrated acceptances of
different analyses. This limit is compared in Fig.~8 with the
limits of
other experiments which searched for bare MMs
 with $g=g_D$ and $\sigma_{cat} <1$ mb [21-28].\par
 Following the same procedure used for MMs, we obtain the 90\% C.L.
global MACRO limit for an isotropic flux of nuclearites
(with masses  $ > 6 \cdot 10^{22}$ GeV/c$^2$, Fig.~6); at
$\beta=2\cdot 10^{-3}$ the
limit is $2.1 \cdot  10^{-16}$ cm$^{-2}$ s$^{-1}$ sr$^{-1}$.
 The MACRO limit for a flux of downgoing
nuclearites is
compared in Fig.~9 with the limits of other experiments [25,29-31].
The Galactic Dark Matter (DM) limit in Fig.~9 was estimated assuming that
$\Phi_{max} = \rho_{DM} v /(2 \pi M)$, where $\rho_{DM} \simeq 10^{-24}$
g/cm$^3$ is the local DM density, and $M$ and $v$ are the mass and
the velocity of nuclearites, respectively.

There is a close connection between searches for Q-balls (type SECS),
nuclearites and magnetic monopoles.
The liquid scintillators, the limited streamer tubes and the nuclear track
detector CR39  are sensitive to SECS. The   limits for MMs
obtained with the scintillators using the PHRASE system and   with
streamer tubes can be applied to SECS.
The global MACRO limit for SECS with electric charge $Z_Q=1$ 
is obtained following the same procedure as for MMs and nuclearites and is
shown in Fig.~7 (curve ``MACRO'').

\vskip 1truecm
\noindent{\bf Acknowledgements.}

We gratefully acknowledge the support of the director and of the staff of the
Laboratori Nazionali del Gran Sasso and the invaluable assistance of the
technical staff of the Institutions participating in the experiment. We thank
the Istituto Nazionale di Fisica Nucleare (INFN), the U.S. Department of
Energy and the U.S. National Science Foundation for their generous support of
the MACRO experiment. We thank INFN, ICTP (Trieste) and World Laboratory
for providing fellowships and grants for non Italian citizens.\par

\begin{figure}
   \begin{center}
      \vspace{-2.cm}
              \epsfysize=13cm
  \epsffile{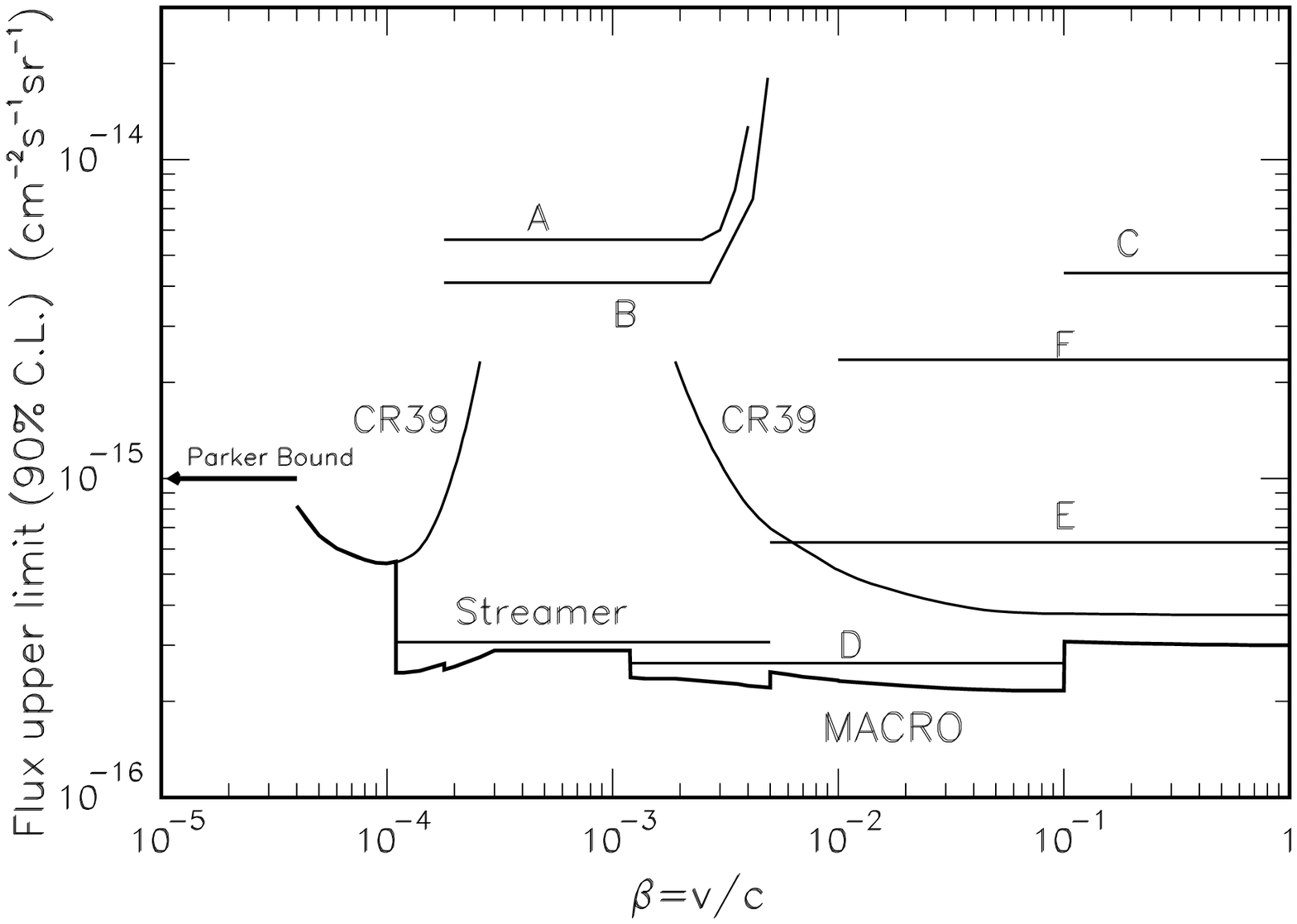}
   \vspace{-1cm}
 \end{center}   

\caption{\small The $90\%$ C.L. upper limits for an isotropic
flux of supermassive magnetic monopoles obtained using the three MACRO subdetectors:
liquid scintillators 
(curves A-D), 
streamer tubes (curve `` Streamer''), nuclear track detectors
(curves ``CR39''); curves E and F were obtained combining the 
three subdetectors together (see text). The bold
line is the present MACRO global limit.}
\end{figure}

\begin{figure}
   \begin{center}
      \vspace{-2.cm}
              \epsfysize=13cm
  \epsffile{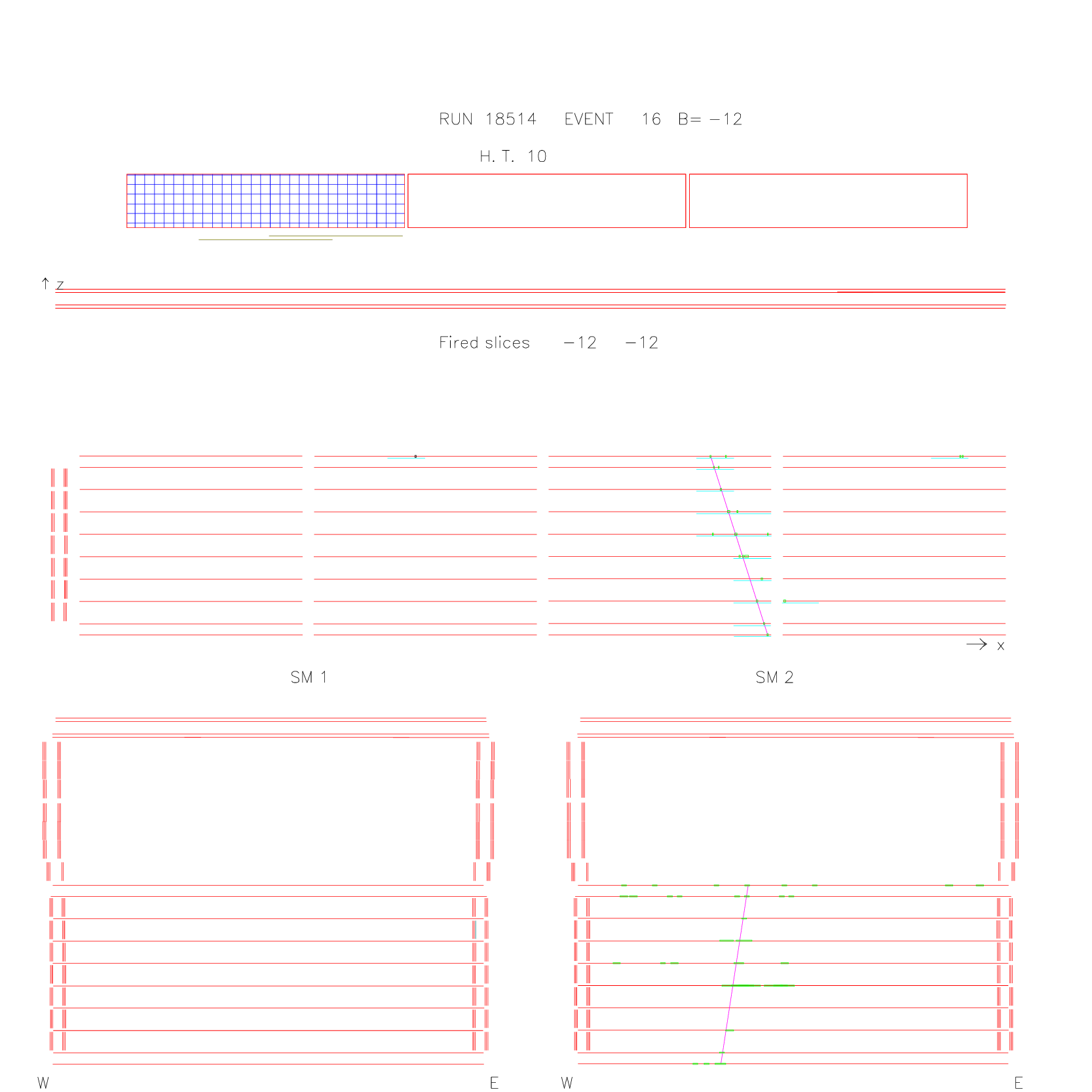}
   \vspace{1cm}
 \end{center}   

\caption{\small Space view (XZ and YZ  in the upper and lower part, 
respectively) of a simulated catalysis event.
The monopole track and the catalysis hits
are clearly distinguishable.}

\end{figure}

\begin{figure}
   \begin{center}
      \vspace{-2.cm}
              \epsfysize=13cm
  \epsffile{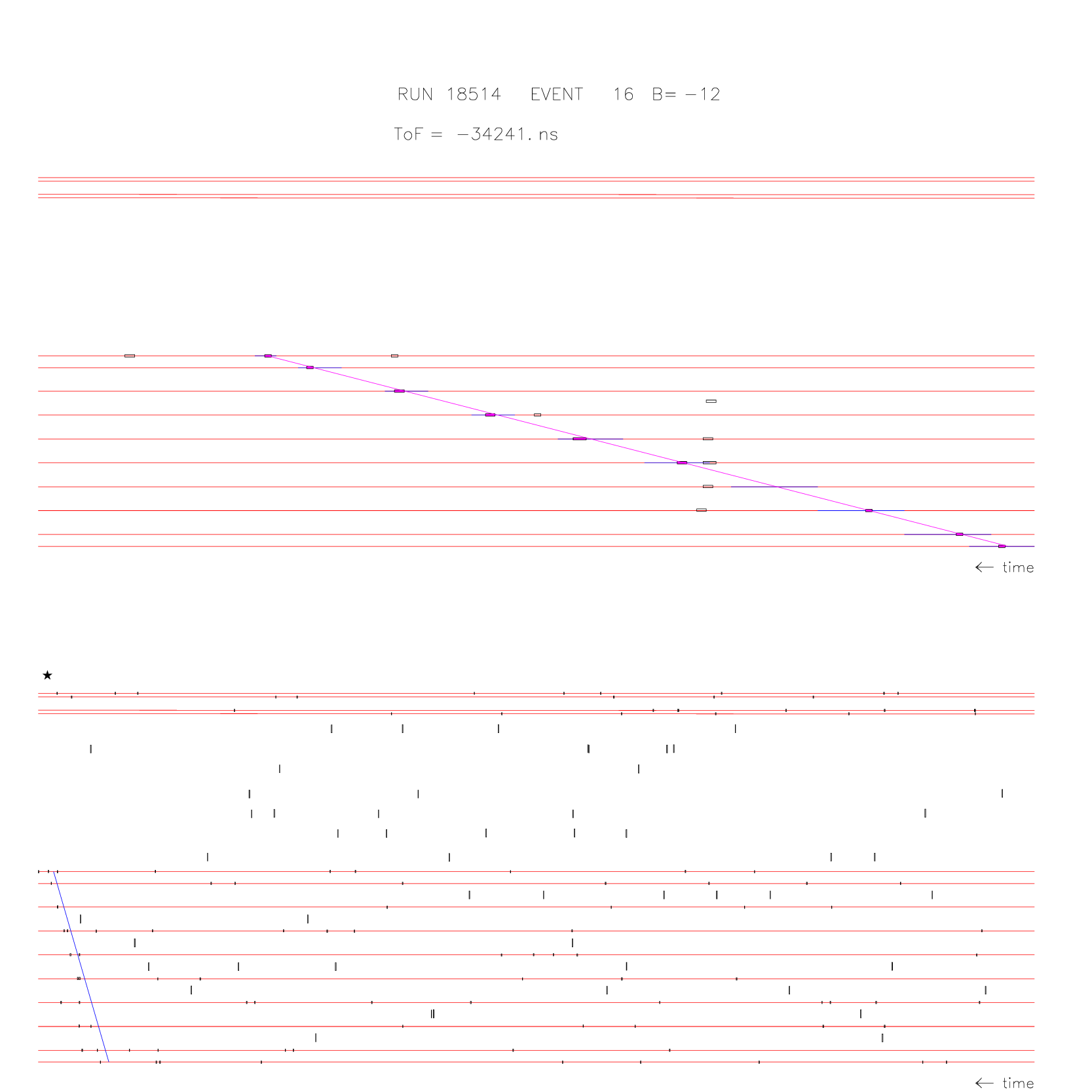}
   \vspace{1cm}
 \end{center}   

\caption{\small Time view of the simulated event shown in Fig.~2. 
In the lower
part the content of the whole $680 \mu s$ QTP memory is plotted; the
upper part shows a magnification of the region around the fired $\beta$
slice. The MM time track is a straight line; the catalysis hits
are grouped in a narrow time window.}


\end{figure}

\begin{figure}
   \begin{center}
              \epsfysize=13cm
  \epsffile{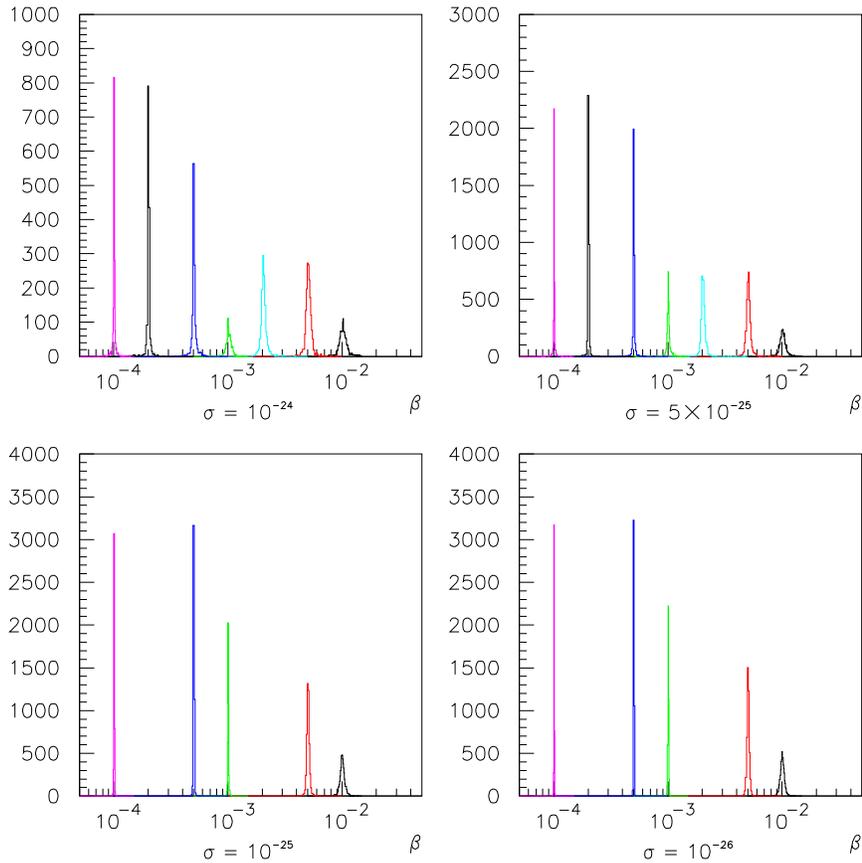}
   \end{center}
\vspace{-1.3cm}
   \caption{\small Distributions of the reconstructed $\beta$  for 
 4 simulated catalysis cross sections. The
distributions are exactly peaked around the input values, which means
that the recontruction procedure (used in the standard streamer analysis), 
gives the correct $\beta$  even in presence
of catalysis hits. }
\end{figure}

\begin{figure}
   \begin{center}
      \vspace{-0.5cm}
              \epsfysize=13cm
     \epsffile{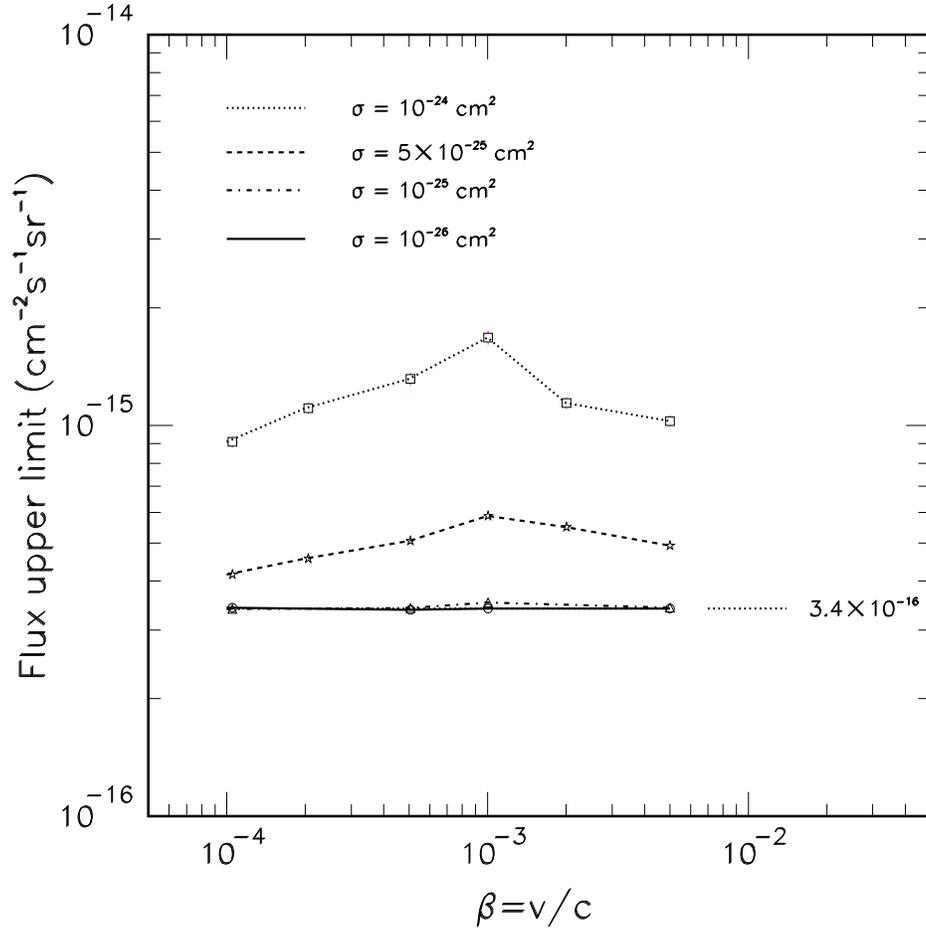}
   \end{center}

\vspace{-1cm}
 
  \caption{\small Flux  limits from streamer tubes 
for 4 values of $\sigma_{cat}$, which is assumed to be the same
 on protons and neutrons: (from above) $\sigma_{cat} =
10^{-24}$, $5\cdot 10^{-25}$, $10^{-25}$ and $10^{-26}~$cm$^2$.
The simulation was performed at fixed $\beta$; the lines
are only a guide to the eye.}
\end{figure}


\begin{figure}
   \begin{center}

      \vspace{-0.5cm}
              \epsfysize=13cm
     \epsffile{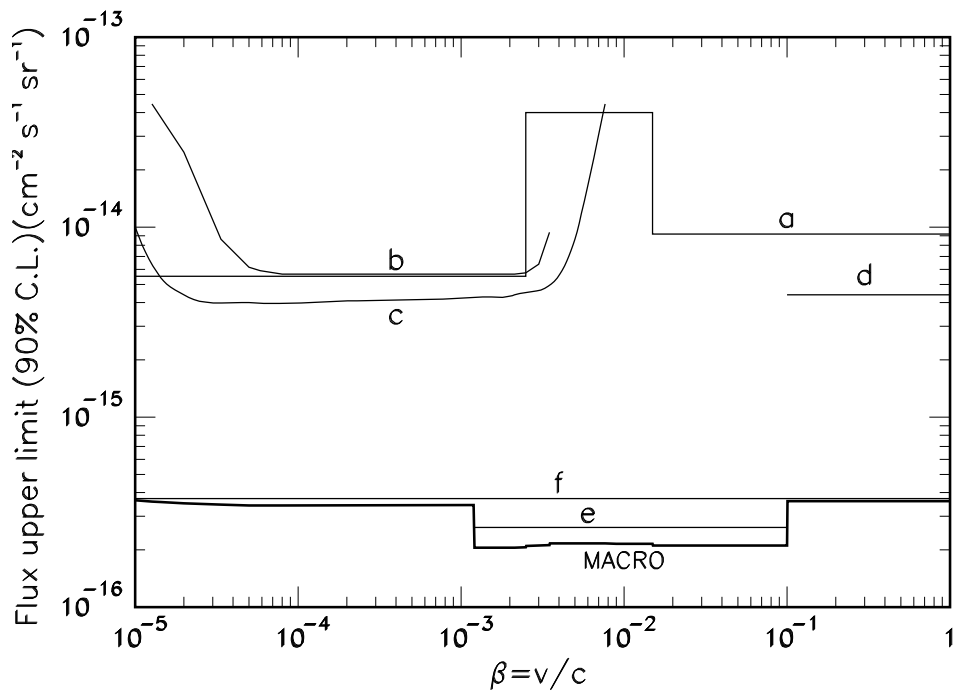}

   \end{center}
\vspace{-1cm}
   \caption{\small The $90\%$ C.L. upper limits for an isotropic flux of
nuclearites obtained using the liquid scintillator (curves ``a'' - ``e'') and
the CR39 nuclear track (curve ``f'') subdetectors; the bold line is the 
MACRO global limit.}
\end{figure}

\begin{figure}
  \begin{center}
              \epsfysize=10cm
\epsffile{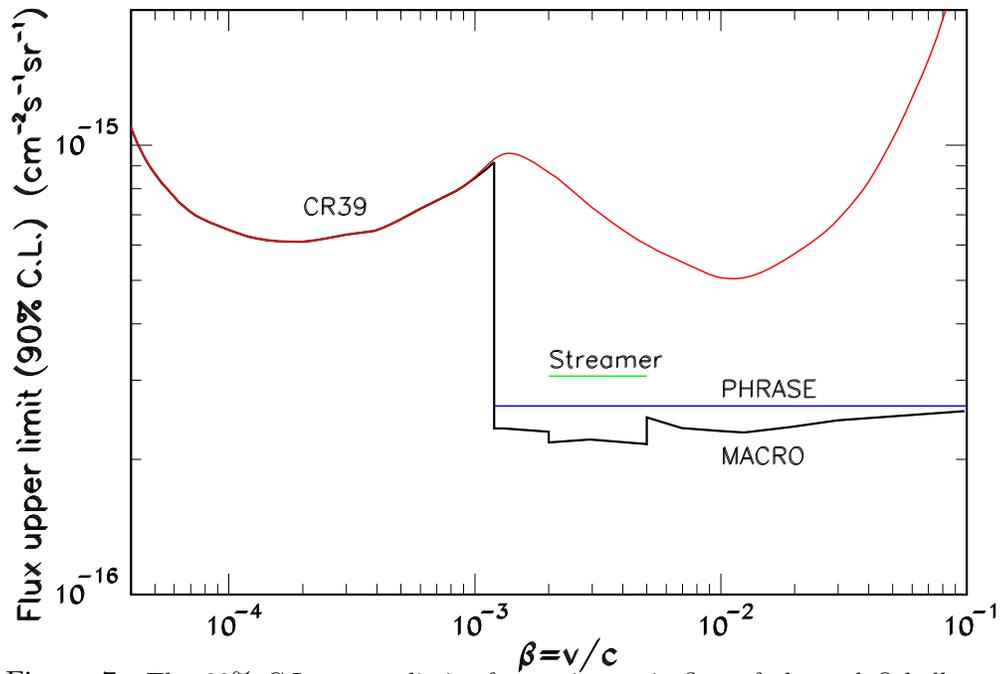}
  \end{center}
\vspace{-1.3cm}
   \caption{\small The $90\%$ C.L. upper limits for an isotropic flux of
charged Q-balls
obtained using the liquid scintillator
(curve ``PHRASE''), 
the CR39 nuclear track detectors (curve ``CR39'') and
 the streamer tubes (curve `` Streamer''). 
The bold line is the MACRO global limit.}
\end{figure}

\begin{figure}
   \begin{center}
              \epsfysize=10cm
  \epsffile{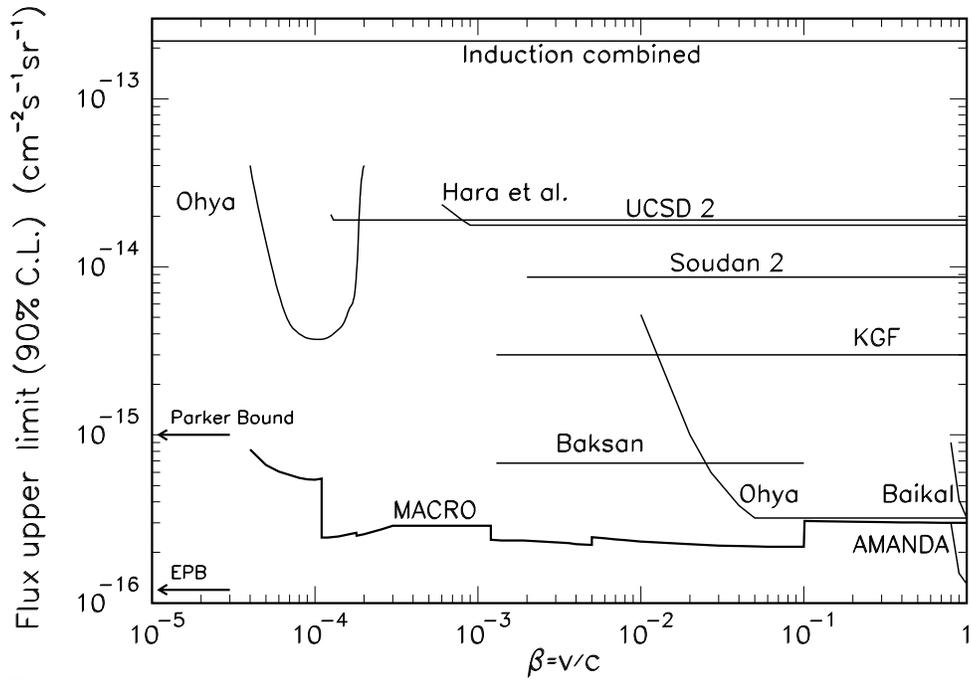}
   \end{center}
\vspace{-1.3cm}
   \caption{\small The global MACRO
$90\%$ C.L. upper limit for an isotropic flux of g=$g_D$
magnetic monopoles extends from $\beta=4\cdot 10^{-5}$ to 1; it 
is compared with the limits obtained by other experiments; at values of
$\beta \simeq 1$ we show the limits from the Baikal and Amanda collaborations.}
\end{figure}

\begin{figure}
   \begin{center}
              \epsfysize=13cm
\epsffile{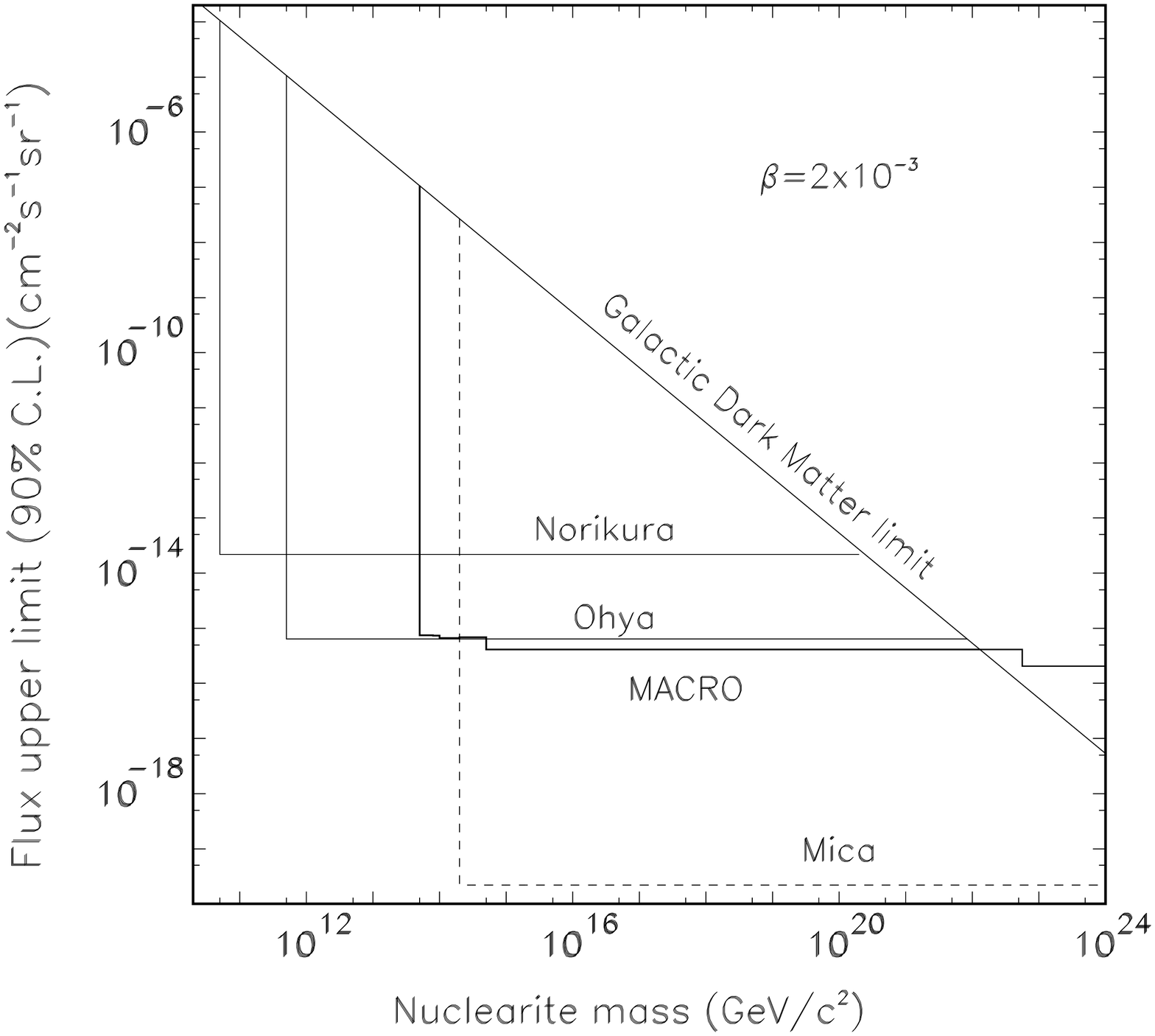}
   \end{center}
\vspace{-1.3cm}
   \caption{\small The global MACRO $90\%$ C.L. upper  flux
limit for nuclearites with $\beta = 2 \cdot 10^{-3}$ at ground level, versus
nuclearite mass, is compared with the limit obtained by other experiments
and with the galactic DM bound. The limit above
$M_N > 6 \cdot 10^{22}$ GeV corresponds to an isotropic flux; for 
$M_N < 6 \cdot 10^{22}$ GeV the limit corresponds to nuclearites reaching 
the detectors only from above.}
\end{figure}

\vskip 1.7cm



\begin{thebibliography}{999}
\parskip -5pt
\bibitem{Preskill79} J.Preskill,  Phys. Rev. Lett. 43 (1979) 1365.
\bibitem{Turner82} M. S. Turner, E. M. Parker and  T. J. Bogdan,
                Phys. Rev. D26 (1982) 1926.
\bibitem{Ahlen93} S. P. Ahlen et al., MACRO Coll., Nucl. Instr. \& Meth.
		  A324 (1993) 337.
\bibitem{Ahlen83} S. P. Ahlen and G. Tarl\'e,  Phys. Rev. D27 (1983) 
			688; G. Giacomelli et al., hep-ex/0005041 (2000).
\bibitem{Batt88} G. Battistoni et al., Nucl. Instr. \& Meth. A270 (1988) 
		    185; G. Battistoni et al., Nucl. Instr. \& Meth. A401 (1997) 309.
\bibitem{Cecco96} S. Cecchini et al., Nuovo. Cim. A109  (1996) 1119.
\bibitem{Witten} E. Witten, Phys. Rev. D30 (1984) 272.
\bibitem{Rujula-Glas84} A. De R\'{u}jula and S. L. Glashow, Nature 312 
			(1984) 734.
\bibitem{Coleman} S. Coleman, Nucl. Phys. B262 (1985) 293.
         
\bibitem{Kusenko} A. Kusenko, Phys. Lett. B404 (1997) 285; Phys. Lett. B405
(1997) 108; 
A. Kusenko and M. Shaposhnikov, Phys. Lett. B417 (1998) 99.
\bibitem{Derkaoui99} J. Derkaoui et al., Astrop. Phys. 10 (1999) 339.
\bibitem{Ambrosio97} M. Ambrosio et al.,  MACRO Coll., Phys. Lett. B406 (1997)
                     249.
\bibitem{Ambrosio92} M. Ambrosio et al.,  MACRO Coll., Astropart. Phys. 
                   1 (1992) 11.
\bibitem{Ambrosio95} M. Ambrosio  et al., MACRO Coll., Astrop. Phys.4 (1995) 33.
\bibitem{Drell}    S. Drell et al., Phys. Rev. Lett. 50 (1983) 644.
\bibitem{Giorgini} M. Giorgini for the MACRO Coll., Nucl. Phys. B (Proc. Suppl)
85 (2000) 227.
\bibitem{Batt97b} G. Battistoni et al., Nucl. Instr. \& Meth. A399 (1997) 244.
\bibitem{Chin71}  S. Chin and A. Kerman, Phys. Rev. Lett. 43 (1971) 1292.
\bibitem{Ambrosio99} M. Ambrosio et al.,  MACRO Coll., hep-ex/9904031,
			Eur. Phys. J. C 13 (2000) 453.
\bibitem{Ouchrif}  D. Bakari et al., hep-ex/0003003, Astrop. Phys. (in press)
(2000).
\bibitem{Bermon90} S. Bermon et al., Phys. Rev. Lett. 64 (1990) 839.
\bibitem{Buckland90} K.N. Buckland et al., Phys. Rev. D41 (1990) 2726.
\bibitem{Thron92} J. L. Thron et al., Phys. Rev. D46  (1992) 4846.
\bibitem{Alexeyev90} E. N. Alexeyev et al., $21^{st}$ ICRC, Adelaide, 
                     vol. 10 (1990) 83.
\bibitem{Orito91} S. Orito et al., Phys. Rev. Lett. 66 (1991) 1951.
\bibitem{Adarkar90} H. Adarkar et al., $21^{st}$ ICRC, Adelaide, vol. 10 (1990)
			95. 
\bibitem{Hara90} T. Hara et al., $21^{st}$ ICRC, Adelaide, vol. 10 (1990) 79.
\bibitem{Amanda} G. Domogatsky for the Baikal Coll. Talk presented at the 
XIX Int. Conf. on
                 Neutrino Physics and Astrophysics, Sudbury, Canada (2000).
\bibitem{Nakamura91} S. Nakamura et al., Phys. Lett. B 263 (1991) 529.
\bibitem{Price88} P.~B.~Price, Phys. Rev. D 38 (1988) 3813.
\bibitem{Ghosh90} D. Ghosh and S. Chatterjea, Europhys. Lett. 12 (1990) 25.

\end{thebibliography}
\end{document}